\documentstyle [12pt,a4p,epsfig,amsmath,multicol]{article}
\begin{document}
\vspace*{0.6cm}

\begin{center} 
{\normalsize\bf Relativistic velocity addition and the relativity of space and time intervals}
\end{center}
\vspace*{0.6cm}
\centerline{\footnotesize J.H.Field}
\baselineskip=13pt
\centerline{\footnotesize\it D\'{e}partement de Physique Nucl\'{e}aire et 
 Corpusculaire, Universit\'{e} de Gen\`{e}ve}
\baselineskip=12pt
\centerline{\footnotesize\it 24, quai Ernest-Ansermet CH-1211Gen\`{e}ve 4. }
\centerline{\footnotesize E-mail: john.field@cern.ch}
\baselineskip=13pt
\vspace*{0.9cm}
\abstract{ A thought experiment first proposed by Sartori is analysed
   using the parallel velocity addition formula of special relativity. The
   spatial and proper-time intervals between some similarly defined spatial
   coincidence events are found to be widely different in different inertial frames.
    This relativity of space and time intervals is quite distinct from the
    well-known time-dilatation and length contraction effects of special relativity.
    Sartori's claimed derivation of the parallel velocity addition formula,
    assuming relativistic time dilatation, based on the thought experiment,
    is shown to be fortuitous.}
\vspace*{0.9cm}
\normalsize\baselineskip=15pt
\setcounter{footnote}{0}
\renewcommand{\thefootnote}{\alph{footnote}}
 \par This paper analyses a thought experiment, that was orginally proposed by
  Sartori~\cite{Sartori}, involving two trains T1 and T2, moving
 at different speeds $v$ and $u$ respectively relative to a fixed platform P.
 It was claimed that
 the parallel velocity addition relation (PVAR) could be derived from the
 experiment by assuming the existence of the relativistic time dilatation effect
 ~\cite{Ein1}. The derivation will be critically examined below; however, the
  main  purpose of the present paper is a different one. {\it Assuming} the PVAR, space
  and time intervals between spatial coincidences, in the direction of motion, of P, T1
  and T2 are calculated
  in their respective rest frames S, S' and S'' and compared. 
\begin{figure}[htbp]
\begin{center}\hspace*{-0.5cm}\mbox{
\epsfysize12.0cm\epsffile{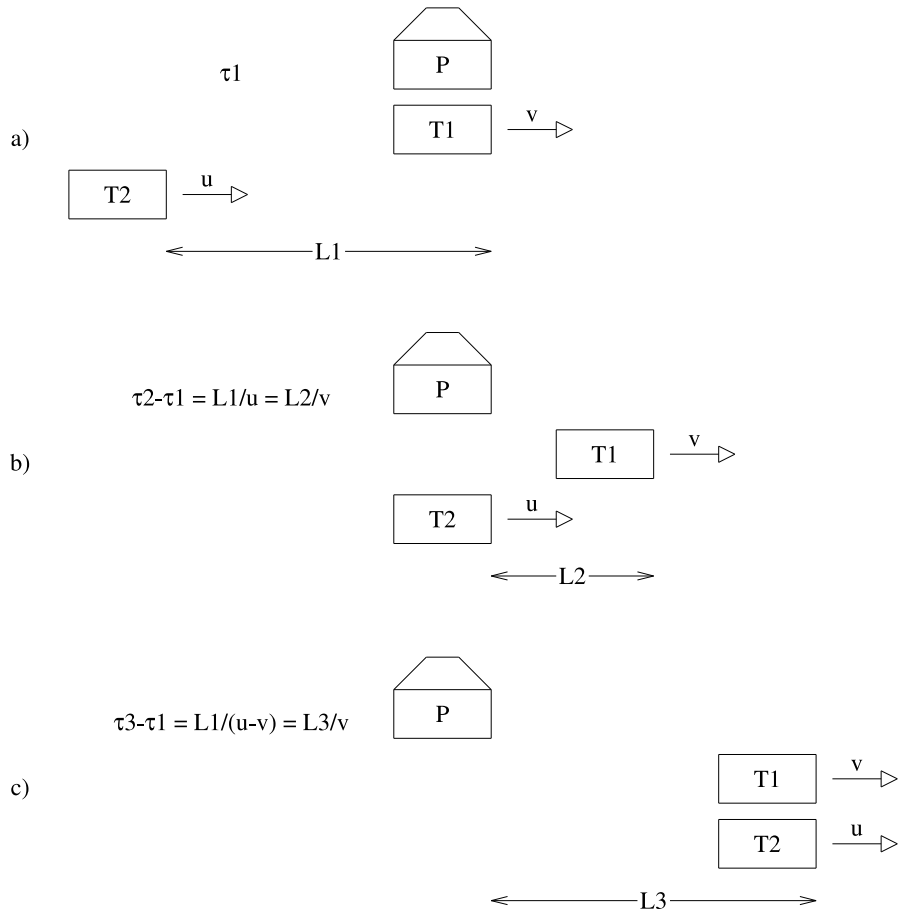}}
\caption{{\em Spatial coincidence events as observed in the rest frame, S, of P.
  a) Event1, T1 opposite P,  b) Event2, T2 opposite P,  c) Event3, T1 opposite T2.
   $u =0.8c$, $v = 0.4c$ }}
\label{fig-fig1}
\end{center}
\end{figure}
\begin{figure}[htbp]
\begin{center}\hspace*{-0.5cm}\mbox{
\epsfysize12.0cm\epsffile{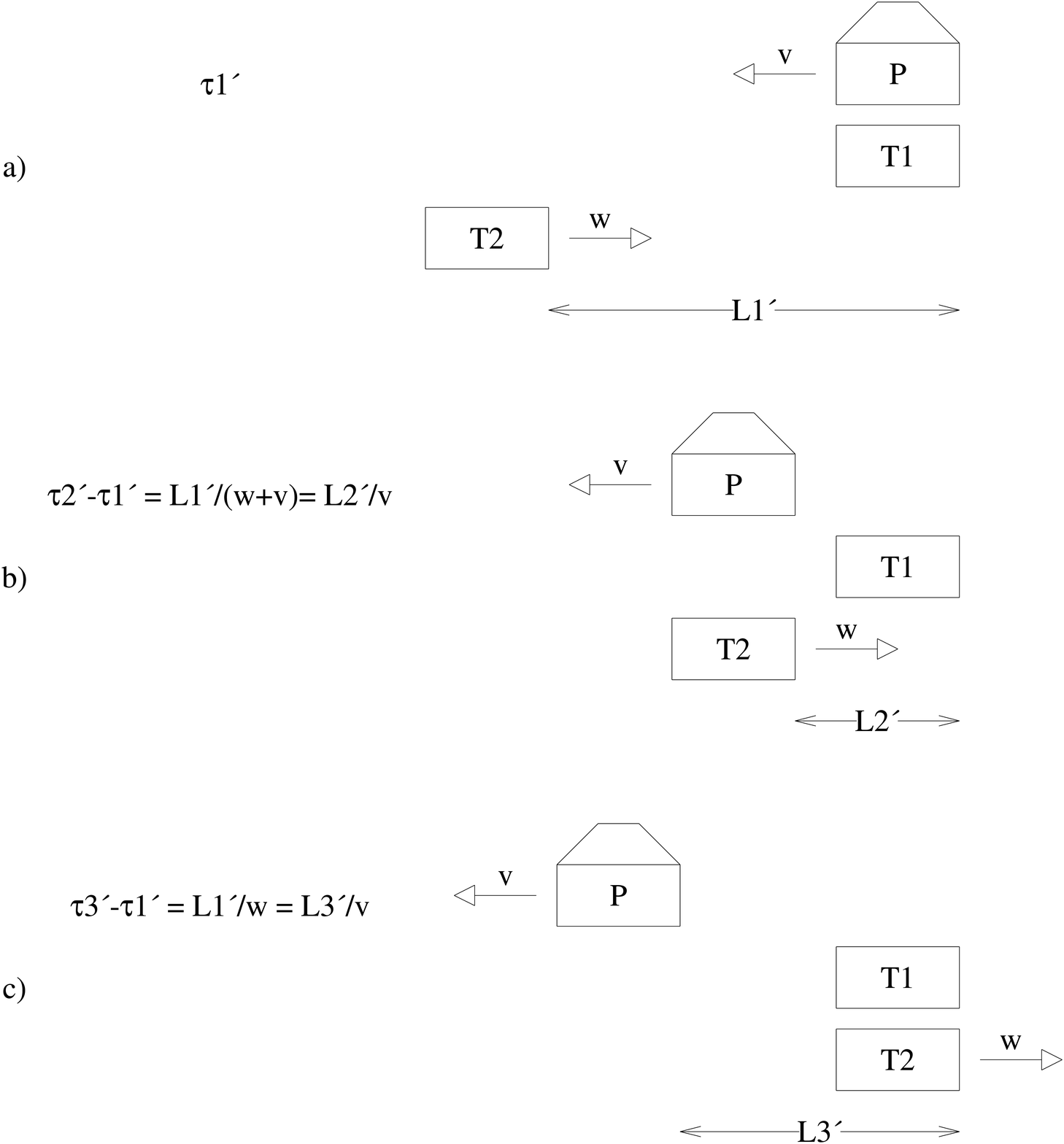}}
\caption{{\em Spatial coincidence events as observed in the rest frame, S', of T1.
  a) Event1, T1 opposite P,  b) Event2, T2 opposite P,  c) Event3, T1 opposite T2.
  $w =0.588c$, $v = 0.4c$  }}
\label{fig-fig2}
\end{center}
\end{figure}
.\begin{figure}[htbp]
\begin{center}\hspace*{-0.5cm}\mbox{
\epsfysize12.0cm\epsffile{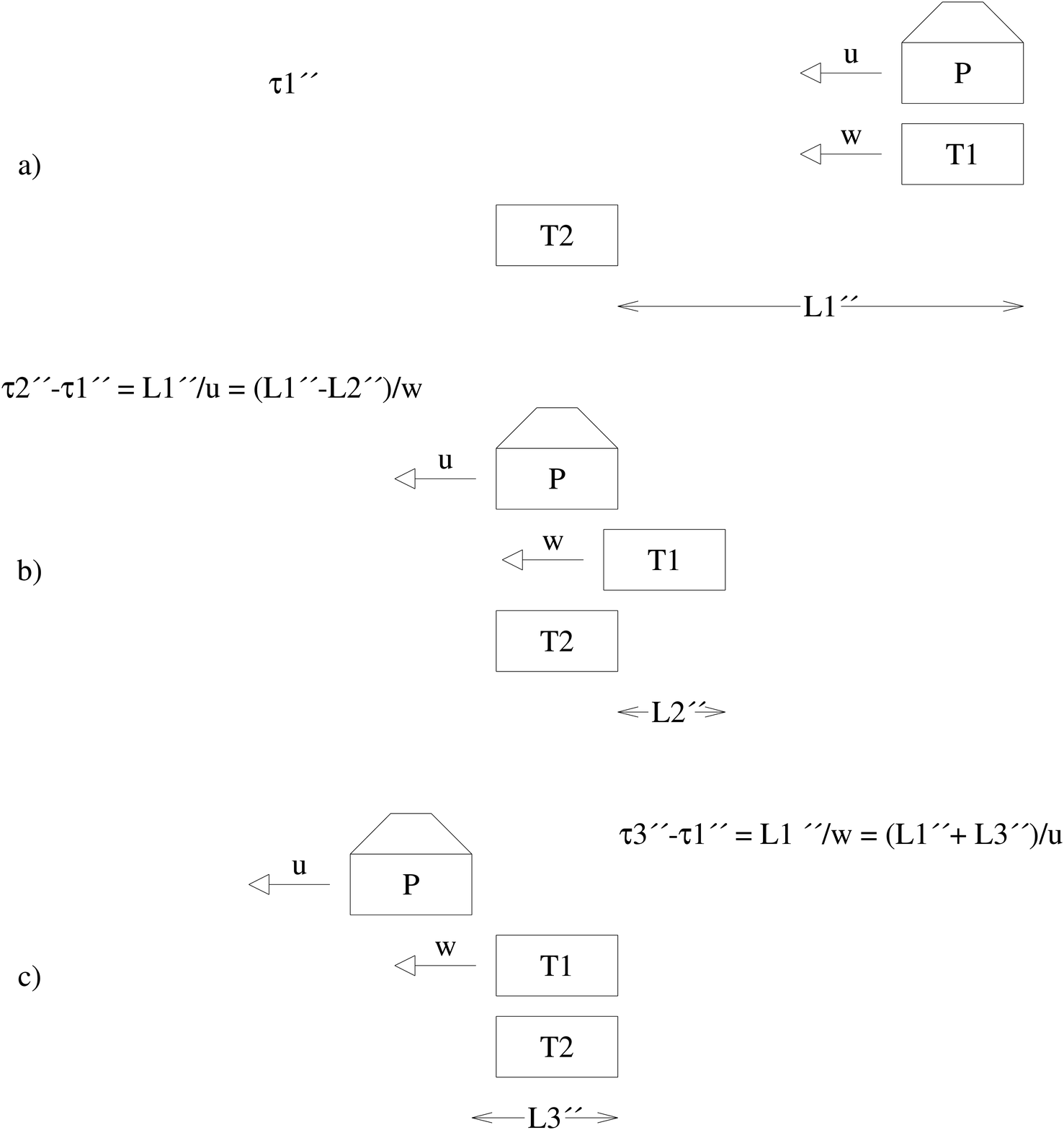}}
\caption{{\em  Spatial coincidence events as observed in the rest frame, S'', of T2.
  a) Event1, T1 opposite P,  b) Event2, T2 opposite P,  c) Event3, T1 opposite T2.
    $u =0.8c$, $w = 0.588 c$}}
\label{fig-fig3}
\end{center}
\end{figure}            
  \par In Fig.1 is shown the experiment as preceived by an observer in the rest frame
 of P. It is assumed that there is an array of synchronised clocks in each frame, each
 of which indicates a common `frame time' (i.e. the proper time of any of the clocks)
 $\tau$, $\tau'$ and $\tau''$ in the frames S, S' and S''. 
At frame time $\tau1 = 0$ the train T1 is opposite the platform (Event1) and T2 is at
 a distance $L1$ from it (Fig1a). At frame time $\tau2$, T2 is opposite the platform (Event2)
  and T1 is at a distance $L2$ from it (Fig1b). Finally, since it is assumed that $u > v$, at
  frame time $\tau3$, T1 and T2 are opposite each other, at a distance $L3$ from P (Event3).
  Figs.2 and 3 show the same
  sequence of events as observed in the rest frames of T1 and T2 respectively.
  The corresponding frame times and distances are $\tau1'$, $\tau2'$,  $\tau3'$; $L1'$, $L2'$
  $L3'$ in the rest frame, S', of T1 and $\tau1''$, $\tau2''$,  $\tau3''$; $L1''$, $L2''$
  $L3''$ in the rest frame, S'', of T2. In Fig.2, T2 and in Fig.3, T1 move at the speed $w$
   given by the PVAR~\cite{Ein1}:
   \begin{equation}
  w = \frac{u -v}{1-\frac{u v}{c^2}}
   \end{equation}
   The geometry of Figs.1-3 gives the relations:
   \begin{eqnarray}
    \tau2- \tau1 & = & \frac{L1}{u},~~L2 = \frac{v L1}{u},~~\tau3- \tau1 = \frac{L1}{u-v},
    ~~L3 = v\frac{L1}{u-v} \\
     \tau2'- \tau1' & = & \frac{L1'}{w+v},~~L2' = \frac{v L1'}{w+v},
     ~~\tau3'- \tau1' = \frac{L1'}{w},~~L3' = v\frac{L1'}{w} \\
  \tau2''- \tau1'' & = & \frac{L1''}{u},~~L2'' = \frac{u-w}{u}L1'',~~\tau3''- \tau1''
     = \frac{L1''}{w},
    ~~L3'' = \frac{u-w}{w}L1''
    \end{eqnarray}
   Eliminating $w$ from (3) and (4) by use of (1), Eqns(2)-(4) yield for the frame times
   and spatial positions of Event2 and Event3 in the frames S, S' and S'', respectively:
      \begin{eqnarray}
    \tau2- \tau1 & = & \frac{L1}{u},~~L2 = \frac{v L1}{u} \\
    \tau3- \tau1 & = & \frac{L1}{u-v},~~L3 = \frac{v L1}{u-v} \\
    \tau2'- \tau1' & = & \frac{1}{u}\frac{(1-\frac{u v}{c^2})}{1-\frac{v^2}{c^2}}L1',
     ~~L2' = \frac{v}{u}\frac{(1-\frac{u v}{c^2})}{1-\frac{v^2}{c^2}}L1' \\
    \tau3'- \tau1' & = & \frac{(1-\frac{u v}{c^2})}{u-v}L1',
     ~~L3' = v\frac{(1-\frac{u v}{c^2})}{u-v}L1' \\
 \tau2''- \tau1'' & = & \frac{L1''}{u},~~L2'' = \frac{v}{u}\frac{(1-\frac{u^2}{c^2})}{1-\frac{u v}{c^2}}L1'' \\
     \tau3''- \tau1'' & = & \frac{(1-\frac{u v}{c^2})}{u-v}L1'',
     ~~L3'' =  v\frac{(1-\frac{u^2}{c^2})}{u-v}L1''
     \end{eqnarray}  
   \par In the above formulae the ratios of length and time intervals in each frame
     are fixed by the relative velocities $u$ and $v$, but the lengths $L1$,  $L1'$ and $L1''$ 
     are arbitary. In order to find the the relation between these lengths it is necessary
 to use the condition that the events corresponding to the spatial coincidences T1-P, T2-P and T1-T2
     are each observed simultaneously in the corresponding inertial frames\footnote{These frames are
      S',S for  T1-P,  S'',S for  T2-P, and  S',S''  for  T1-T2.}. 
     \par Inspection on Figs.1-3  shows that there are nine distinct observations of spatial
    coincidence events in the problem: T1-P, T2-P and T1-T2  as observed in the frames S, S' and S''.
    It is convenient to introduce the following notation to specify these different observations:
    \par Event1~~~~~~~\underline{P(S)-T1(S')}~~~~~~~~\underline{P(S)'-T1(S')'}~~~~~~~~P(S)''-T1(S')''
    \newline
    \par Event2~~~~~~~\underline{P(S)-T2(S'')}~~~~~~~~P(S)'-T2(S'')'~~~~~~~~\underline{P(S)''-T2(S'')''}
     \newline
  \par Event3~~~~~~~T1(S')-T2(S'')~~~~~~~\underline{T1(S')'-T2(S'')'}~~~~~~\underline{T1(S')''-T2(S'')''}
     \newline 
  \par Here, for example, P(S) shows that the proper frame of P is S, and P is observed in S,
       whereas P(S)' indicates that P is observed in the frame S'. The other symbols are defined
      in a similar fashion. The underlined observations are {\it reciprocal} ones of the same
       spatial coincidence event. In each case, one object is at rest, and the other in motion 
       with the same absolute velocity. These observations have the important property that
       they are {\it mutually simultaneous}\footnote{To be contrasted with simultaneous
        events in the {\it same} inertial frame, e.g. those specified by the same reading
        of two stationary synchronised clocks at different positions in the frame.} in the proper
       frames of both objects. This
       property was used by Einstein~\cite{Ein1} to define synchronous clocks in two inertial
       frames at the instant of spatial coincidence of the origins of the frames,
       by setting both frame times to zero at this instant. This procedure has been
       called `system external synchronisation' by Mansouri and Sexl~\cite{MS}.
        In order to mutually synchronise local clocks at the positions of P, T1 and T2,  
        system external synchronisation may be applied at any two of the three 
        pairs of reciprocal observations. For example {P(S)-T1(S') and P(S)'-T1(S')'
        in S and S' respectively may be used to set $\tau1 = \tau1' = 0$. This was
         done in Sartori's analysis of the problem. The mutually simultaneous
        observations P(S)-T2(S'') and {P(S)''-T2(S'')'' may be then used to set 
         $\tau2'' = \tau2$. However a clearer understanding of the problem
            is obtained by performing the analysis of the problem without
        introducing synchronised clocks. Such an analysis is now presented.
         \par Since Event1 is mutually simultaneous in the frames S and S', it follows
          that a necessary condition that two other events with times $\tau$ and $\tau'$ 
         are mutually simultaneous is that
         \begin{equation}
          \tau - \tau' =  \tau1 - \tau1' \equiv \Delta \tau(\rm{S,S'})
          \end{equation}
        Similarly, since Event2 is mutually simultaneous in the frames S and S'',
        two other events in these frames are mutually simultaneous provided that
         \begin{equation}
          \tau - \tau'' =  \tau2 - \tau2'' \equiv \Delta \tau(\rm{S,S''})
          \end{equation}
        It follows from (11) and (12) that events with times $\tau'$ and $\tau''$
        are mutually simultaneous in the frames S' and S'', provided that:
    \begin{equation}
          \tau' - \tau'' = \Delta \tau(\rm{S,S''}) - \Delta \tau(\rm{S,S'})
          = \tau2-\tau2''-( \tau1 - \tau1')
          \end{equation}
     Since the Event3 {\it is} mutually simultaneous in the frames S' and S'' it follows
     that:
      \begin{equation}
          \tau3' - \tau3''  = \tau2-\tau2''-(\tau1 - \tau1')
      \end{equation}
    This equation may be rearranged to give:
      \begin{equation}
          \tau3' - \tau1'  = \tau2- \tau1 +\tau3'' -\tau1''-(\tau2'' - \tau1'')
      \end{equation}
     where $\tau1''$ has been added to and subtracted from the right side.
      Combining (15) with Eqns(2),(3) and (4) gives
       \begin{equation}
         \frac{L1'}{w} = \frac{L1}{u}+  \frac{L1''}{w} - \frac{L1''}{u}
      \end{equation}
     solving this equation for $L1''$:
     \begin{equation}
         L1'' = \frac{L1'-(w/u)L1}{1-w/u}
      \end{equation}
       This equation holds for all values of $w/u$. Setting $w = 0$ then gives
     \begin{equation}
         L1'' =  L1'
      \end{equation}
     Using (18) to eliminate $L1''$ from (16) gives 
  \begin{equation}
          L1 = L1' =   L1''
   \end{equation}
   In view of (19) the following relations may be derived from (5)-(10):
       \begin{eqnarray}
    \frac{\Delta\tau21'}{\Delta\tau21}& = & \frac{1-\frac{u v}{c^2}}{1-\frac{v^2}{c^2}} = \frac{L2'}{L2},~~
  \frac{\Delta\tau31'}{\Delta\tau31} = 1-\frac{u v}{c^2} =  \frac{L3'}{L3 } \\
     \frac{\Delta\tau21'' }{\Delta\tau21}& = & 1,~~ \frac{L2''}{L2} =\frac{1-\frac{u^2}{c^2}}{1-\frac{u v}{c^2}},
   ~~\frac{\Delta\tau31''}{\Delta\tau31} = 1-\frac{u v}{c^2},
   ~~\frac{L3''}{L3} = 1-\frac{u^2}{c^2}
          \end{eqnarray}
 where, for example, $\Delta\tau21 \equiv \tau2-\tau1$.  
     Setting $u = c$ in (20) and (21) gives:
       \begin{eqnarray}
    \frac{\Delta\tau21'}{\Delta\tau21}& = & \frac{1}{1+\frac{v}{c}} = \frac{L2'}{L2},~~
  \frac{\Delta\tau31'}{\Delta\tau31} = 1-\frac{v}{c} =  \frac{L3'}{L3 } \\
     \frac{\Delta\tau21'' }{\Delta\tau21}& = & 1,~~ \frac{L2''}{L2} = \frac{L3''}{L3} = 0,
       ~~\frac{\Delta\tau31''}{\Delta\tau31} = 1-\frac{v}{c} 
          \end{eqnarray} 
     The relativistic contraction of $L2'$ and $L2''$ relative to $L2$, and
     $L3'$ and $L3''$ relative to  $L3$,
      is evident on comparing Fig.1 with Figs.2 and 3, where $u = 0.8c,~v =0.4c~w = .588c$.
      The reciprocal observations of the coincidence events are mutually simultaneous:
     \[ \frac{\Delta \tau21''}{\Delta \tau21} = \frac{\Delta \tau31''}{\Delta \tau31'} = 1 \]
       These relations exemplify the `Measurement Reciprocity Postulate ~\cite{JHF1}
         that reciprocal measurements in two inertial frames yield identical results
        --in this case equal time intervals in the frames in which the measurements are performed 
       \par For the coincidences P(S)'-T2(S'')' for Event2 and T1(S')-T2(S'') for Event3, where both
      objects are in motion in the frame of observation, (20) and(21) give:
   \[ \frac{\Delta \tau21''}{\Delta \tau21'} = \frac{1-\frac{v^2}{c^2}}{1-\frac{u v}{c^2}} \ne 1~~~
        \frac{\Delta \tau31''}{\Delta \tau31} = 1-\frac{u v}{c^2}  \ne 1 \]
    so that that non-reciprocal observations of spatial coincidences (where, in one frame, 
    both objects are in motion) are found to be not mutually simultaneous.
     Note that the 
      relativity of lengths and times apparent in (20) and (21) is quite distinct from
      the time dilatation (TD) and length contraction effects of conventional
      special relativity.
      All times considered here are frame times recorded by stationary clocks 
      in each frame, so that the Lorentz transformation, 
      that relates a proper time recorded by stationary clock to the apparent time
     of uniformly moving one, plays no role, except insofar as the PVAR
     is a necessary consequence of the LT~\cite{Ein1}.
      
    \par Sartori's derivation of the PVAR~\cite{Sartori}, on the basis of the thought
      experiment just discussed, is now considered. It is based on the relations, readily
     derived from Eqns(2) and (3) above:
          \begin{eqnarray}
         \Delta\tau31 & = & \frac{u}{u-v}\Delta\tau21 \\
          \Delta\tau31' & = & \frac{w+v}{w}\Delta\tau21'
      \end{eqnarray}
      In order to derive the PVAR from these equations Sartori assumes that the
      frame time intervals $\Delta\tau21$ and $\Delta\tau21'$ and 
      $\Delta\tau31$ and $\Delta\tau31'$ are connected by the
      TD relations~\cite{Ein1}:
           \begin{eqnarray}
           \Delta\tau21' & = & \gamma \Delta\tau21 \\
           \Delta\tau31  & = & \gamma \Delta\tau31'
          \end{eqnarray}
         where $\gamma \equiv \sqrt{1-(v/c)^2}$.  
        The argument given by Sartori to justify (26) is that: `Since Event1 and Event2 
       occur at the same place in S the interval $\tau2-\tau1$' (in the notation 
       of the present paper)`is a proper time interval'. For (27) the argument is:
       `Similarly since Event1 and Event3 occur at the same place in S',
        the interval $\tau3'-\tau1'$ is a proper time interval'.          
      These statements are correct, but since the times $\tau$, $\tau'$, are those recorded
      by a synchronised clock at {\it any} position in the frames S, S' the fact that the time
      interval $\tau2-\tau1$ is defined by spatial coincidences with P (P-T1 at the
        beginning of the interval and P-T2 at the end) has no special physical
       significance. Similarly, for the interval $\tau3' -\tau1'$, where the limits
      of the interval are defined by the spatial coincidences T1-P and T1-T2, the exact
        correspondence of this interval with the proper time interval of a clock at rest at
        the position of T1 is in no way different, given the possible existence
        of a synchronised clock indicating the frame time at any position in the frame,
        to a frame-time interval defined
        by spatial coincidences at different positions. 
         \par The experiment defined by the TD relation (26) is one in which a local clock situated
        at P is observed from the frame S', relative to which it is moving to the left with
        velocity $v$ (Fig.2) If the frame-time interval $\Delta\tau21'$ in Eqn(26) is that between the
        events P-T1 and P-T2 in S', it  is not correct to substitute $\Delta\tau21$, the frame
        time interval $\tau2-\tau1$ in Fig.1 in the TD relation. In fact (26) and (27) should be written as
             \begin{eqnarray}
           \Delta\tau21' & = & \gamma \Delta t21 \\
           \Delta\tau31  & = & \gamma \Delta t31'
          \end{eqnarray}
        where $\Delta t21 \equiv t_2-t_1$ is the apparent time interval of a
       moving `slowed down' clock at the position of P as viewed from S' and $\Delta t31' \equiv t_3'-t_1'$
  is the apparent time interval of a moving `slowed down' clock at the position of T1 as viewed from
     S. In the thought experiment analysed above no such observations of moving clocks are performed,
      so the correct TD formulae (28) and (29) have no possible relevance to the problem. Indeed, Sartori's
      formulae (26) and (27) are incorrect.
        Eqns(20) and (21) give
          \begin{eqnarray}
      \frac{\Delta \tau21'}{\Delta \tau21} & = & \frac{1- \frac{uv}{c^2}}{1-\frac{v^2}{c^2}} \ne \gamma \\
   \frac{\Delta \tau31}{\Delta \tau31'} & = & \frac{1}{1-\frac{uv}{c^2}} \ne \gamma
    \end{eqnarray}
    In contradiction to Eqns(26) and (27).
      \par The hypothesis underlying (26) and (27) --the substitution of frame times defined by the coincidence
       events shown in Figs.1 and 2 
     into the TD relation-- may be excluded on the grounds that it contradicts the assumed initial
      conditions of the thought experiment. Consider the experiment reciprocal to that
      described by the TD relation (28) i.e. a local clock at the position of T1
      is viewed from the frame S. Although there is no local spatial coincidence
      to define the time interval $\tau2'-\tau1'$ in Fig.2, this interval is
      the same as that which would be recorded by a local clock at T1 at the instant of
     the P-T2 coincidence, since $\tau2'$ is the time of this coincidence as
     recorded by {\it all} synchronised clocks in S'. Sartori's ansatz then gives,
     for this reciprocal experiment, the relation
      \begin{equation}
       \Delta \tau21 = \gamma \Delta \tau21'
       \end{equation}
      Similar consideration of reciprocal experiments, gives, with Sartori's hypothesis, the relation
      reciprocal to (27): 
    \begin{equation}
       \Delta \tau31' = \gamma \Delta \tau31
       \end{equation}
 
       Combining (26) and (32) gives 
      \begin{equation}
        \Delta\tau21'  =  \gamma  \Delta\tau21 =  \gamma^2  \Delta\tau21'
   \end{equation}
      It then follows that $\gamma^2 = 1$, $v = 0$, contradicting the initial hypothesis
      $v \ne 0$ of the thought experiment. The same conclusion follows on combining (27) and (33).
       \par The apparent time intervals $\Delta t21$ and $\Delta t31'$ as observed in the
        frames S' and S (28) and (29) are also frame time intervals
       $\Delta t21 = \Delta \tilde{\tau}21 \equiv  \tilde{\tau}2 - \tau1$ and
        $\Delta t31' = \Delta \tilde{\tau}31' \equiv  \tilde{\tau}3'- \tau1'$ seen by observers
        in S and S' respectively. The spatial configurations
        in these frames given by (28) and (29) corresponding to the values of
      $\Delta \tau21'$ and $\Delta \tau31$ in Figs.2 and 1 respectively are shown in Figs.4 and 5.
       Combining (28) or (29) with (5)-(8) and (19) gives:
        \begin{eqnarray}
          \Delta \tilde{\tau}21 & = & \frac{1-\frac{uv}{c^2}}{\gamma u (1-\frac{v^2}{c^2})}L1,~~~\Delta \tau21 = \frac{L1}{u } \\
   \Delta\tilde{\tau}31' & = & \frac{L1}{\gamma(u-v)},~~~\Delta \tau31' = \frac{L1}{w}
         \end{eqnarray}
      With $u = 0.8c$, $v = 0.4c$, $w = 0.588c$ and $\gamma = 1.09$, as in the figures
   \begin{eqnarray}
        \Delta \tilde{\tau}21 & = & 0.93 L1/c < \Delta \tau21 = 1.25 L1/c \nonumber \\
         \Delta\tilde{\tau}31' & = & 2.3  L1/c > ~\Delta \tau31' = 1.7 L1/c \nonumber
    \end{eqnarray}
      In Fig.4 $\tilde{\tau}2$ is less than $\tau2$, whereas in Fig.5  $\tilde{\tau}3'$ is greater
       than  $\tau3'$.
      \par Eqns(24) and (25) can be combined to give the relation
        \begin{equation}
       \frac{\Delta\tau31}{\Delta\tau31'} \frac{\Delta\tau21'}{\Delta\tau21}(1-\frac{v}{u})(1+\frac{v}{w}) = 1
  \end{equation}
       Substituting the ratios $\Delta\tau31/\Delta\tau31'$ and $\Delta\tau21'/\Delta\tau21$
   from (20) gives
      \begin{equation}
       (1-\frac{v}{u})(1+\frac{v}{w}) = (1-\frac{v^2}{c^2})
  \end{equation}
     which, when solved for $w$ in terms of $u$ and $v$ gives the PVAR (1).
     The same result is given by substituting the ratios  $\Delta\tau3/\Delta\tau3'$ and
      $\Delta\tau2'/\Delta\tau2$
     from Sartori's relations (26) and (27). 
     Making the equally (in)valid substitution of frame times for apparent times
     to give the reciprocal TD relations of (32) and (33), and substituting
     these times into Eqn(37) gives
     \begin{equation}
      (1-\frac{v^2}{c^2})(1-\frac{v}{u})(1+\frac{v}{w}) = 1
  \end{equation}
     which is derived by similar hypotheses to that used to derive (38), but does not yield the PVAR.
      In conclusion, Sartori's
     `derivation' of the correct PVAR (1) on the basis of the the assumed (but incorrect) equations (26) and (27)
      is fortuitous. 
\begin{figure}[htbp]
\begin{center}\hspace*{-0.5cm}\mbox{
\epsfysize7.0cm\epsffile{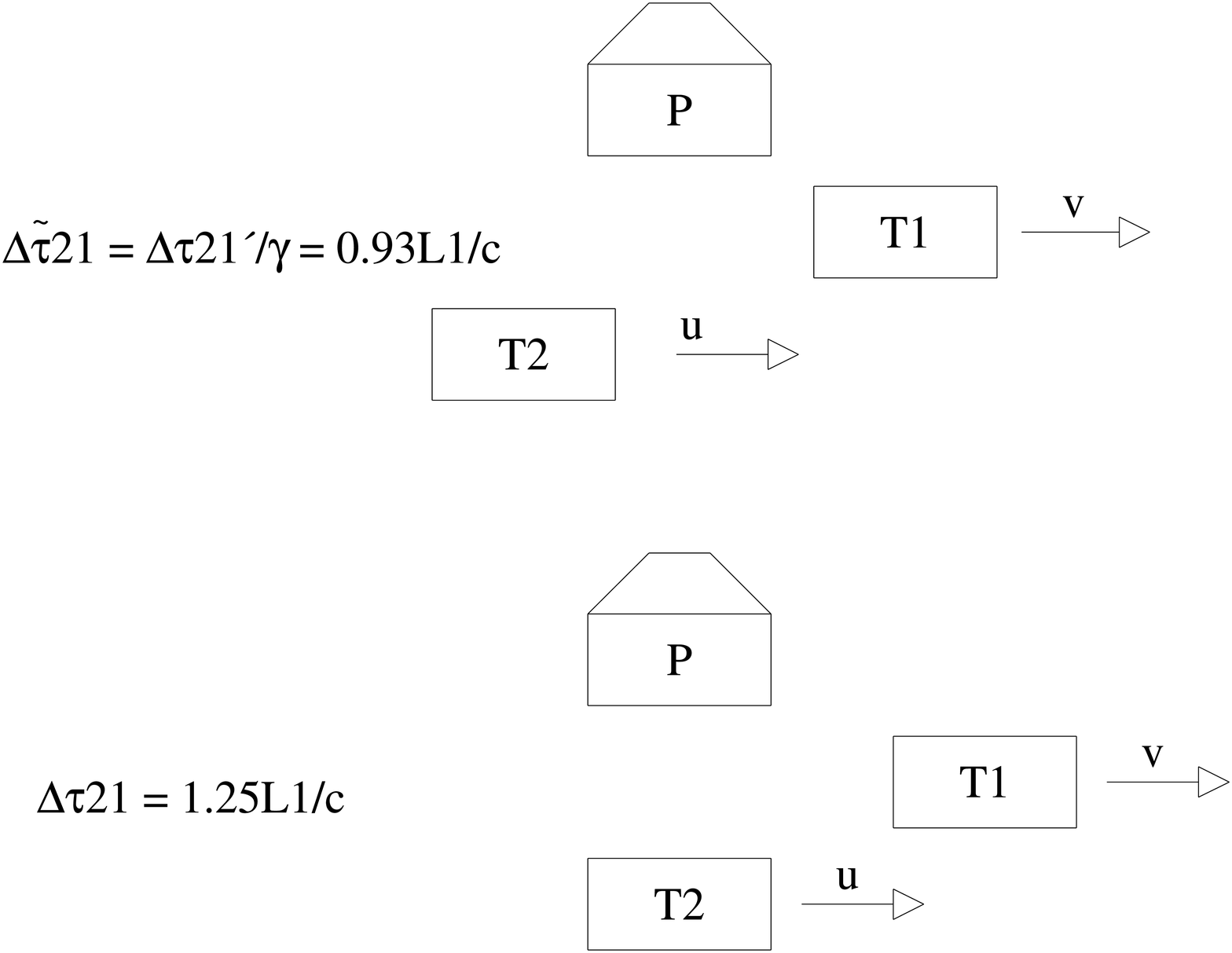}}
\caption{{\em Spatial coincidence events as observed in the rest frame, S, of P.
  Comparison of configurations at times $\tau 2$ and $\tilde{\tau}2$. The event at time $\tilde{\tau}2$
   is related via the TD relation (28) with the event at time $\tau 2'$ in S'.}}
\label{fig-fig4}
\end{center}
\end{figure}
\begin{figure}[htbp]
\begin{center}\hspace*{-0.5cm}\mbox{
\epsfysize7.0cm\epsffile{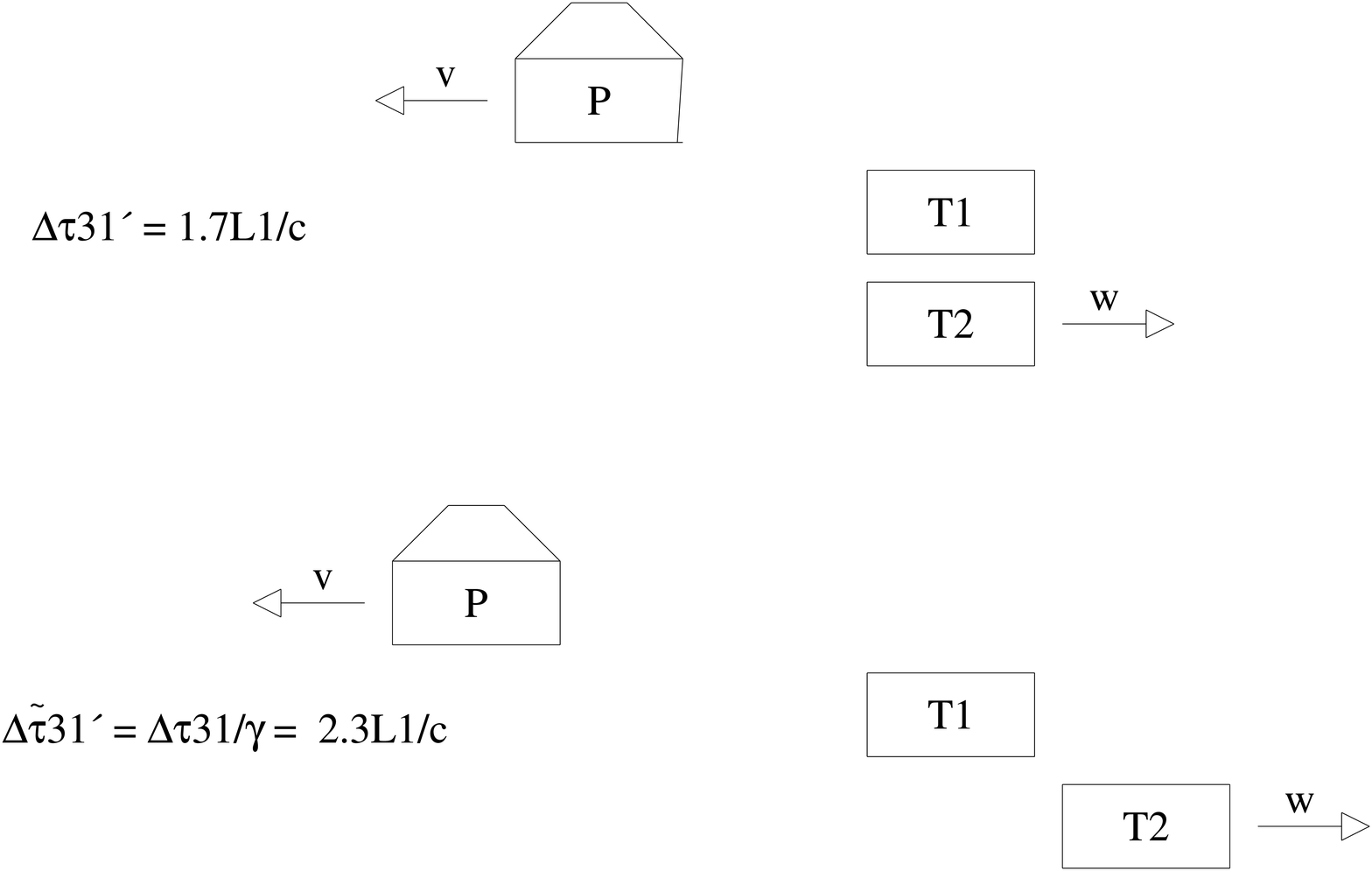}}
\caption{{\em Spatial coincidence events as observed in the rest frame, S', of T1.
 Comparison of configurations at times $\tau 3'$ and $\tilde{\tau}3'$. The event at time $\tilde{\tau}3'$
   is related via the TD relation (29) with the event at time $\tau3$ in S.}}
\label{fig-fig5}
\end{center}
\end{figure}
\par
\par {\bf Acknowledgement}
 
\par  I thank two anonymous refereees of another journal for critical comments that prompted me
    to improve the clarity of the presentation of two earlier versions of this paper. 
 \par {\bf Added Note}
 \par Some conclusions of the present paper are flawed by important errors both in calculation and of
  concept. The correct analysis of Sartori's thought experiment is given in a more recent 
   paper by the present author~\cite{JHFPRSTE}. The known calculational and conceptual mistakes that
   exist in the analysis of the present paper are explained in this Added Note.
 \par The argument concerning mutually simultaneous events,. leading to Eqn(17), is logically correct.
  The aim of this calculation is to find the relation of $L1'$ and $L1''$ to $L1$, assuming the latter
   distance to be a fixed initial parameter of the problem, and that  $L1'$ and $L1''$ are observed
 in the frames S', S'' moving with speeds $v$, $u$, respectively, relative to the frame S in which 
   $L1$ is specified. The separations $L1'$ and $L1''$ must then have (if any) a similar velocity
   dependence: $L1' = L(L1,v)$, $L1'' = L(L1,u)$ where $L1 \equiv L(L1,0)$. Thus Eqn(17) should
   be written:
     \begin{equation}
         L(L1,u) = \frac{L(L1,v)-(w/u)L(L1,0)}{1-w/u}
      \end{equation}
      If $w = 0$ then $u = v$ so that (40) does not give, as claimed, the relation $L1'' = L1'$ but instead
      the trivial identity:
    \begin{equation}
     L(L1,v) \equiv L(L1,v)
     \end{equation}
    \par However Eqn(19): $L1 =  L1' = L1''$  , is correct. To show this, introduce $x$-coordinate axes
      in the frames S and S'' parallel to the direction of motion of the trains, with origins aligned with
     the position of P in Fig.1a at $\tau = \tau1$. The world lines of T2 in S'' and S respectively are then
     \[ x''({\rm T}2) = - L1'' ({\rm all}~\tau''),~~~   x({\rm T}2) = u(\tau-\tau1)- L1 \]
       corresponding to the space Lorentz transformation equation:
     \begin{equation}
     x''({\rm T}2)+  L1''  = \gamma_u[ x({\rm T}2)+ L1 - u(\tau-\tau1)] = 0 
     \end{equation}   
      where $\gamma_u \equiv1/\sqrt{1-(u/c)^2}$. The interval $ L1''$ is a fixed constant and
   $L1 \equiv - x({\rm T}2,\tau = \tau1)$ is also a constant, which is independent of $u$, reflecting the choice of spatial
   coordinate system in S. Therefore Eqn(42) must hold for 
   all values of $u$; in particular, it holds when $u = 0$, $\gamma_u = 1$ and $x' \rightarrow x''$, giving
     \begin{equation}
     x''({\rm T}2)+  L1''  = x''({\rm T}2)+ L1
     \end{equation}
    so that  
   \begin{equation}
     L1''  =  L1
     \end{equation}
 A similar argument for the frames S and S', considering the configuration of P and T1 at time $\tau = \tau1-L1/v$, 
 shows that:
    \begin{equation}
     L1'  =  L1
     \end{equation}
 so that Eqn(19) is verified.
    \par The formulae (2)-(4), as derived from the geometry of Figs.1-3, are correct. However, if the
     configurations of Figs.2 and 3 are to
    correctly represent (as claimed by Sartori) observations in the frames S' and S'', respectively, of the coincidence
    events defined in the frame S in Fig.1, then the PVAR of Eqn(1) must be replaced by the Relative Velocity
     Addition Relation (RVAR) Eqn(5.19) of Ref.~\cite{JHFPRSTE}, which corresponds, in the notation for
    velocities in Figs.1-3, to:
   \begin{equation}
   w = \gamma_v(u-v)~~~({\rm Fig.}2)
   \end{equation}
   \begin{equation}
   w = \gamma_u(u-v)~~~({\rm Fig.}3)
   \end{equation}
     while the speed $v$ of P in Fig.2 is replaced by $v\gamma_v$ and $u$ in Fig.3 by  $u\gamma_u$ (see Figs. 3 and 4
     of  Ref.~\cite{JHFPRSTE}). Thus also $v$ in Eqn(3) should be replaced by $v\gamma_v$,  $u$ in Eqn(4) by  $u\gamma_u$. 
     The velocity $w$ in Eqn(3) is replaced by that in Eqn(46), that in Eqn(4) is by that in Eqn(47).
     The times and separations in Eqns(5)-(10) are then replaced by those presented in the first row  of
      Table 2 in  Ref.~\cite{JHFPRSTE}. It is found that $L2 =  L2' = L2''$  and $L3 =  L3' = L3''$ so that there
      is no special `length contraction' effect as claimed in Eqns(20) and (21).
     \par Also, in contradiction to what is shown in Figs.4 and 5, events at times connected by the TD relation
      ($\Delta\tau21' = \gamma \Delta \tau21$ in Fig.4, $\Delta\tau31 = \gamma \Delta \tau31'$ in Fig.5) correspond to the 
        same spatial coincidence events (P-T2 in the primary experiment in Fig.4, T1-T2 in the
        reciprocal experiment in Fig.5). The distinction between $\Delta\tilde{\tau}21$ and
       $\Delta\tau21$ and between  $\Delta\tilde{\tau}31'$ and $\Delta\tau31'$ claimed in these figures and the accompanying
       calculations is therefore illusory.
     \par What are actually shown in Figs.2 and 3 ---configurations reciprocal to those shown in Fig.1 and related to them
     by the PVAR of Eqn(1)--- are configuration of related but {\it physically independent} space-time experiments in
     which the
     initial velocity parameters $u$ and $v$ of Fig.1 are replaced by $v$ and $w$ in Fig.2 and by $u$ and $w$ in Fig.3.
      In the nomenclature introduced in Refs.~\cite{JHFSTP3,JHFPRSTE} what are shown in Figs.2 and 3 are `base frame'
      configurations in S' and S'' that are reciprocal to the one in S shown in Fig.1. Physically distinct TD effects
       (see Eqns(7.1)-(7.3) of Ref.~\cite{JHFPRSTE}) occur in the three different experiments. 
         \par The claim that Sartori's equations (26) and (27) above are erroneous because incompatible with Eqns(30) 
      and (31) is wrong since, as explained above,  these equations (taken from Eqns(20) and (21) respectively) are 
   incorrect. However, the essential critique of Sartori's derivation of the PVAR ---the irrelevance, given the existence
      of synchronised clocks at different, and in principle, arbitary, positions in an inertial frame
      of a physical local clock in specifying the proper time interval in the TD relation, and the
     use of TD relations in an experiment and its reciprocal (which  are physically independent) as though they
     related observations of the same events in different frames in the primary experiment--- remains valid. In particular,
     the demonstration of the self-contradictory nature, as shown in Eqn(34), of any attempt to associate the TD effect
     of the reciprocal experiment with observations of the same events in different frames in the primary experiment
     is not affected by any of the previous calculational or conceptual errors in the paper. Sartori's `derivation' of
     the PVAR therefore remains fortuitous.

\end{document}